\documentclass[aip, jcp, reprint, floatfix]{revtex4-1}
\usepackage{amsmath,amsbsy,amssymb,amsfonts}
\usepackage{graphicx, color}
\usepackage{empheq}
\newcommand{\Matr}[1]{\boldsymbol{\mathcal{\hat{#1}}}}
\newcommand{\crn}[1]{\hat{#1}^{\dagger}}
\newcommand{\anh}[1]{\hat{#1}}
\renewcommand{\vec}[1]{\pmb{#1}}

\graphicspath{{PictForPaper/}}

\begin{document}
\date{\today}
\title{Fast transport and relaxation of vibrational energy in polymer chains}
\author{Arkady A. Kurnosov}
\author{Igor V. Rubtsov}
\author{Alexander L. Burin}
\email[Author to whom correspondence should be addressed. Email: ]{aburin@tulane.edu}
\affiliation{Department of Chemistry, Tulane University, New Orleans, Louisiana 70118, USA}
\begin{abstract}
We investigate \textit{ballistic} vibrational energy transport through optical phonon band in oligomeric chains in the presence of decoherence. An exact solution is obtained for the excitation density in the space-time representation in the continuous limit and this solution is used to characterize the energy transport time and intensity. Three transport mechanisms are identified such as \textit{ballistic}, \textit{diffusive} and \textit{directed diffusive} regimes, occurring at different distances and time ranges. The crossover between the two \textit{diffusive} regimes is continuous, while the switch between the \textit{ballistic} and \textit{diffusive} mechanisms  occurs in discontinuous manner in accord with the recent experimental data on energy transport in perfluoroalkanes.      
\end{abstract}
\maketitle
\section{\label{Intro} Introduction}
The transport on a molecular scale is usually associated with coherent and incoherent mechanisms. The coherent transport is characterized by energy and phase conservation and the absence of environment disturbance. In the event of semi-classical motion with no under-barrier tunneling involved such transport it can be also called \textit{ballistic} transport. Both coherent and incoherent regimes and transition between them have been a long term focus for the scientific community. Such transitions were studied for electronic, \cite{joachim2005molecular, yu2005thermal, wang2006carbon} exciton \cite{haken1973exactly, schwarzer1975coherent} and more recently vibrational energy transport.\cite{leitner2001vibrational, yu2003vibrational, schwarzer2004intramolecular, schroder2009molecular, wang2007ultrafast, lin2012constant,  rubtsova2013ballistic}

In the \textit{ballistic} regime, the energy is transferred via a free-propagating vibrational wavepacket, formed by vibrational states delocalized over the whole transport region; such transport can be very efficient and fast.\cite{henry2008high, shen2010polyethylene} Such transport can be expected in polymer molecules composed of nearly identical units, where the normal modes are formed by superposition of many excited monomer states; otherwise the \textit{diffusive} mechanism is mostly prevalent.

The \textit{diffusive} energy transport is a result of intramolecular \textit{vibrational energy redistribution} (IVR), which involves energy hopping between vibrational states. Diffusive transport is expected to occur in molecules lacking periodic structure \cite{tesar2013theoretical} with the normal modes localized at the length comparable to the inter-atomic distance. A single IVR event, serving as a driving force of \textit{dif\-fu\-sive} energy transport, is characterized by a change of three or more quantum numbers of the involved spatially overlapping vibrational modes, which requires anharmonic coupling of these modes.

In oligomers normal vibrational modes can be substantially delocalized because of the translational symmetry. Therefore one can expect observing of \textit{ballistic} transport in such systems. Indeed, this transport has been observed in bridged azulene-anthracene compounds,\cite{schwarzer2004intramolecular} polyethylene glycol oligomers,\cite{lin2012constant} alkanes\cite{wang2007ultrafast} and perfluoroalkanes \cite{rubtsova2013ballistic} and the theory describing it has been suggested. \cite{segal2003thermal, benderskii2011propagating, benderskii2013vibrational} 

In a recent paper \cite{rubtsova2014temperature} the energy transport via highly ordered perfluoroalkane chains was studied by the relaxation-assisted two-dimensional infrared method. \cite{rubtsov2009relaxation, kurochkin2007relaxation} The \textit{bal\-lis\-tic} transport with the speed of 385~m/s was found and the transport time showed strong temperature dependence.\cite{rubtsova2014temperature} To interpret the observations, the authors developed a simple model describing the \textit{bal\-lis\-tic} transport and its decoherence caused by dynamic fluctuations of the environment. The strong temperature dependence has been interpreted assuming that the transition from the \textit{bal\-lis\-tic} regime to the \textit{dif\-fu\-sive} regime takes place with the increase of the temperature. Since this transition is of a general interest here we investigate it in greater detail. 

In this study we describe the model of \textit{bal\-lis\-tic} transport and decoherence in greater detail and derive its analytical solution in the space-time representation extending the earlier results of Haken and Strobl\cite{haken1973exactly} and Schwartzer\cite{schwarzer1975coherent} to the practical regime of interest.\cite{rubtsova2013ballistic, henry2008high, shen2010polyethylene, tesar2013theoretical, segal2003thermal, benderskii2011propagating, benderskii2013vibrational, rubtsova2014temperature} To the best of our knowledge this solution has never been reported yet. Then we consider asymptotic limits that describe different  transport mechanisms, including \textit{bal\-lis\-tic}, \textit{dif\-fu\-sive} and  \textit{directed diffusive} regimes, as well as the crossovers between them. 

\section{\label{Model} The model}
Consider a polymer chain composed of $N$ identical monomers with the only one relevant vibrational mode (for example C-F stretching or F-F bending modes in perfluoroalkanes) on each site forming an optical phonon band. The Hamiltonian can be expressed as
\begin{equation}
\Matr{H} = \frac{\hbar}{2}\sum\limits_{m = 1}^N\omega_m \crn{b}_m\anh{b}_m + \frac{\hbar\Delta}{2}\sum\limits_{m = 1}^{N - 1}\left(\crn{b}_m\anh{b}_{m + 1} + \anh{b}_m\crn{b}_{m + 1}\right)
\end{equation}
where $\omega_m$ is the average vibration frequency on the $m$-th site and $\Delta$ is the coupling of the neighboring sites. We assume that the average site frequencies are the same for each site, 
\begin{equation}\label{Eq:assumption1}
<\omega_m> = \omega_0
\end{equation} 
and their fluctuations are delta-correlated:
\begin{equation}\label{Eq:assumption2}
<\delta\omega_m(t_1)\delta\omega_n(t_2)> = \frac{W}{2}\delta_{m n}\delta(t_1 - t_2)
\end{equation}
where $\delta_{m n}$ is the Kronecker delta.
The assumption of Eq.(\ref{Eq:assumption1}) is justified by the high  ordering of perfluoroalkane chains. The assumption of Eq.(\ref{Eq:assumption2}) is the standard approximation, which treats the site frequencies as uncorrelated, while introducing the decoherence rate of $W/2$ for each site. \citep{haken1973exactly, skinner1986pure, bulatov1998effect}  

 We consider the low temperature case, $k_{B}T \ll \hbar\omega$, so the thermal excitations of vibrational states can be neglected and the only excitation in the chain is caused by the external laser pulse. The time evolution of this excitation can be described in terms of the density matrix $\rho_{mn} = |m><n|$, where $|m>$ denotes the state with excitation on the $m$-th monomer. The density matrix satisfies the quantum \textit{Liouville - Bloch equation} \cite{haken1973exactly, schwarzer1975coherent, bulatov1998effect}

\begin{equation}
\label{Eq:Liouv}
\frac{\partial \vec{\rho}}{\partial t} = -\frac{i}{\hbar}\left[\Matr{H}, \vec{\rho}\right] - \Matr{W}\vec{\rho} - \gamma\vec{\rho}
\end{equation}

\begin{equation}
\left(\Matr{W}\vec{\rho}\right)_{m n} = W(1 - \delta_{m n})\rho_{m n}
\end{equation}
where $W$ is the decoherence for all off-diagonal elements and $\gamma$ stands for the pure dissipation rate (e. g. relaxation to the solvent). We assume that the decoherence rate is much larger than the dissipation rate. This agrees with the experimental data analysis\cite{rubtsova2014temperature} and common sense expectation, because the decoherence comes from the energy fluctuations while the dissipation requires a real transition. 

The probability to observe excitation on the $n$-th site, which is referred to as a signal intensity, is given by  the diagonal density matrix element $\rho_{nn}$. To characterize the excitation transport we consider the simplest model of infinite chain, $N = \infty$, and assume that initially only a single site with $n=0$ is populated.  Finally we are to solve the following system of equations:
\begin{multline}
\label{Eq:SetUp}
\frac{\partial \rho_{m n}}{\partial t} = -i\Delta\left\{\rho_{m-1 n} + \rho_{m+1 n} - \rho_{m n-1} -\rho_{m n+1}\right\}\\
-(W+\gamma)\rho_{m n} + W\delta_{m n}\rho_{m n}
\end{multline}
\begin{equation}
\label{Eq:InCond}
\rho_{m n}(0) = \delta_{0 m}\delta_{0 n}
\end{equation}

\section{\label{Solution} Solution}
The solution of Eq.(\ref{Eq:SetUp}) for the diagonal elements of the density matrix  can be obtained in the exact form in the continuous limit, $N \gg 1$. This limit is equivalent to evaluating of the \textit{inverse Fourier transform} of the solution in the momentum representation obtained in Ref.\cite{haken1973exactly}.  Importantly this solution can be evaluated in the analytical form in space-time representation. 
The probability of finding the excitation on the $n$-th site is given by the expression 
\begin{multline}
\label{Eq:LiouvillAnswer}
P_n(t) = 
e^{-(W + \gamma) t}\Bigg\{\mathrm{J}_n^2(2\Delta t) +
 \frac{W}{4\Delta}\Bigg[\mathrm{I}_0\left(W\sqrt{t^2 - \frac{n^2}{4\Delta^2}}\right)  \\
+ \mathrm{L}_0\left(W\sqrt{t^2 - \frac{n^2}{4\Delta^2}}\right)\Bigg]\theta\left(t^2 - \frac{n^2}{4\Delta^2}\right)\Bigg\}
\end{multline} 
where $\mathrm{J}_n$ and $\mathrm{I}_0$ are the $n$-th order and zero-order Bessel functions, respectively, and $\mathrm{L}_0$ is the zero-order Struve function. \cite{abramowitz1972handbook} The derivation of the Eq. (\ref{Eq:LiouvillAnswer}) is given below.   
\subsection{\label{GenSol} General Approach}
Applying \textit{Fourier transform}, $\tilde{\rho}(p, k; t) = \sum\rho_{m n}\exp\{i a(p n - k m)\}$, to Eq. (\ref{Eq:SetUp}) with respect to  both indices, $m$ and $n$, we obtain 
\begin{multline}  
\label{Eq:FourierTime}
\dot{\tilde{\rho}}(p, k; t) = -i 2\Delta\left\{\cos\left(p a\right) - \cos\left(k a\right)\right\}\tilde{\rho}(p, k) \\
- (W + \gamma)\tilde{\rho}(k, p) + W \tilde{P}(p - k)
\end{multline}
where
\begin{equation}\label{Eq:G}
\tilde{P}(q) = \sum\limits_n e^{i a q n}\rho_{n n}
\end{equation}
is the \textit{Fourier transform} of the site-diagonal density matrix characterizing transport of the excitation density and $a$ is the average distance between two adjacent monomers. 

If one can manage to find $\tilde{P}(q)$, then the diagonal components of the density matrix can be found as \textit{inverse Fourier transform}:
\begin{equation}
\label{Eq:G2P}
\rho_{n n}(t) = \left(\frac{a}{2\pi}\right)^2\int\limits_{-\pi/a}^{+\pi/a} d p\int\limits_{-\pi/a}^{+\pi/a} d k e^{-i (p - k) a n}\tilde{P}(p - k; t)
\end{equation}
After applying \textit{Laplace transform}  with respect to time to Eq. (\ref{Eq:FourierTime})
\begin{equation}\label{Eq:LTdef}
\tilde{\rho}(z) = \mathcal{L}_z\left[\tilde{\rho}(z)\right] = \int\limits_0^{+\infty}d t e^{-z t}\tilde{\rho}(t)
\end{equation}
one can represent the density matrix in terms of its diagonal part as 
\begin{equation}\label{Eq:LapTrans}
\tilde{\rho}(p, k; z) = \frac{1 + W \tilde{P}(p - k; z)}{z + i 2\Delta\left[\cos\left(p a\right) - \cos\left(k a\right)\right] + W + \gamma}
\end{equation}
where $2\Delta \cos(pa)$ describes the spectrum of optical phonons within the band. The group velocity for the specific wave vector $p$ is given by $2a\Delta \sin(pa)$. The maximum velocity corresponds to $p = \pi/(2a)$ and is given by $v_{max} = 2a\Delta$. We will show  below that this is the actual velocity of the \textit{ballistic} transport. 
Using the definition of $\tilde{P}(q)$ 
\begin{equation}\label{Eq:G(q)}
\tilde{P}(q) = \frac{a}{2\pi}\int\limits_{-\pi/a}^{+\pi/a}dp\int\limits_{-\pi/a}^{+\pi/a}dk \delta(p - k - q)\tilde{\rho}(p, k)
\end{equation}
and the identity
\begin{multline}
\int\limits_{-\pi/a}^{+\pi/a}d p\int\limits_{-\pi/a}^{+\pi/a}d k\frac{\delta(p - k - q)}{z + W + \gamma + i 2\Delta\left[\cos(p a) - \cos(k a)\right]}\\
 =\frac{2\pi}{a}\frac{1}{\sqrt{[z + W + \gamma]^2 + \left[4\Delta\sin(q a/2)\right]^2}}
\end{multline} 
and solving Eq. (\ref{Eq:LapTrans}) for $\tilde{P}$ we get
\begin{equation}
\label{Eq:Gz}
\tilde{P}(q; z) = \frac{1}{\sqrt{[z + W + \gamma]^2 + \left[4\Delta\sin(q a/2)\right]^2} - W}
\end{equation}
To obtain the probability of finding excitation on the $n$-th site, $P_{n}(t) = \rho_{n n}(t)$, we  need to apply the \textit{inverse Fourier transform} and the \textit{inverse Laplace transform}  to Eq. (\ref{Eq:Gz}). Before doing that we will split Eq.(\ref{Eq:Gz}) in two components, $\tilde{P} = \tilde{P}_B + \tilde{P}_D$, defined below in Eqs. (\ref{Eq:Gball}) and (\ref{Eq:Gdiff}) and discuss their physical meaning. The first component 
\begin{equation}\label{Eq:Gball}
\tilde{P}_B(q) = \frac{1}{\sqrt{[z + W + \gamma]^2 + \left[4\Delta\sin(q a/2)\right]^2}}
\end{equation} 
is responsible for the \textit{ballistic} transport. Indeed, if we consider transport to large distance so that $q$ is small and $\sin(q a/2) \simeq q a/2$, then  Eq. (\ref{Eq:Gball})  corresponds to the running wave-packet with the group velocity $v = 2a\Delta$ and dumping rate $W + \gamma$.  
The \textit{diffusive} part takes the form
\begin{equation}\label{Eq:Gdiff}
\tilde{P}_D(q; z) = \frac{W \tilde{P}_B(q; z)}{\sqrt{[z + W + \gamma]^2 + \left[4\Delta\sin(q a/2)\right]^2} - W}
\end{equation}

In the case of  energy conservation and long distance - long time limit it can be expressed in the form of a diffusion pole  $\tilde{P}_D(q; z) \sim 1/(z + 2 D q^2)$, where $D = a^2\Delta/W$ is a diffusion coefficient.
Next we will consider \textit{ballistic} and \textit{diffusive} components of the solution separately.

\subsection{\label{Coherent} Ballistic Transport}
Based on the property of the \textit{inverse Laplace transform} one can see that the \textit{ballistic} component is given by $\exp\{-(W+\gamma)t\}\rho^0_{nn}(t)$, where $\rho^0_{mn}(t)$ is a solution of Eq. (\ref{Eq:SetUp}) in a purely coherent case, $W = \gamma = 0$. $\rho^0_{mn}(t)$ can be found as a tensor product of wave-functions, $\vec\rho = |\psi><\psi|$, wich are given by

\begin{equation}
\psi_n(t) = e^{i\frac{n\pi}{2}}\mathrm{J}_n(2\Delta t)
\end{equation}

The probability, $\rho_{nn}^{0}(t)$, to find the excitation on the $n$-th site (in agreement with Ref. \cite{schwarzer1975coherent}) is given by 
\begin{equation}\label{Eq:P0}
P_n^B(t) = e^{-(W + \gamma)t}\mathrm{J}_n^2(2\Delta t)
\end{equation}


\subsection{\label{Diffusion}Diffusive Transport}
For the \textit{diffusive} component the double integral in Eq. (\ref{Eq:G2P}) can be reduced to
\begin{equation}\label{Eq:DoubleInverse}
P^D_{n}(t) =  \mathcal{L}^{-1}_t\left[\frac{a}{4\pi}\int\limits_{-2\pi/a}^{+2\pi/a}d q \tilde{P}_D(q; z)e^{-i q a n}\right]
\end{equation}
where $\mathcal{L}^{-1}_t$ denotes the $z \longrightarrow t$ \textit{inverse Laplace transform}. We will expand Eq. (\ref{Eq:Gdiff}) into a series with respect to the number of scattering events on the stochastic random potential $\delta\omega(t)$ associated with the decoherence and express the solution as 
\begin{equation}
\label{Eq:SolveWithSum}
P^D_{n}(t) = \sum\limits_{k = 1}^{\infty}\rho_{n n}^{k}(t)
\end{equation} 
where
\begin{multline}
\label{Eq:Rawrhom}
\rho_{n n}^{k}(t) = \frac{a}{4\pi}\frac{W^{k}}{2\pi i}\underset{\delta\rightarrow 0}{\lim}\int\limits_{-i\infty}^{+i\infty}d z\int\limits_{-2\pi/a}^{+2\pi/a}d q e^{-i q a n}e^{(z + \delta)t}\\
\times\frac{1}{\left[(z + \delta + W + \gamma )^2 + \left(4\Delta\sin(q a/2)\right)^2\right]^{\frac{k-1}{2} + 1}}
\end{multline} 
For long distances, $n\gg 1$, which are the target of our consideration, one can expand $\sin(q a/2)\simeq q a/2$ and set all integration limits to infinity, which corresponds to the exact limit of the continuous model. 

Applying those assumptions and introducing new variables, $z_{\pm} = z \pm i2\Delta q + W + \gamma$, we obtain
\begin{equation}
\rho^{k}_{n n}(t) = \frac{W^{k}}{4\Delta}\mathcal{U}\left(t - \frac{n}{2\Delta}\right)\mathcal{U}\left(t + \frac{n}{2\Delta}\right)e^{-(W + \gamma)t}
\end{equation}
where
\begin{multline}\label{Eq:IntTau}
\mathcal{U}(\tau) = \frac{1}{2\pi i}\underset{\delta\rightarrow 0}{\lim}\int\limits_{-i\infty}^{+i\infty}d z\frac{e^{\frac{1}{2}(z + \delta)\tau}}{(z + \delta)^{\frac{k-1}{2} + 1}}\\
= \mathcal{L}^{-1}_{\tau/2}\left[\frac{1}{z^{\frac{k-1}{2} + 1}}\right] = \frac{\left(\frac{\tau}{2}\right)^\frac{k-1}{2}}{\Gamma(\frac{k-1}{2} + 1)}\theta(\tau) 
\end{multline}
and $\theta(\tau)$ is a Heaviside step function.  
 
Introducing $x = W\sqrt{t^2 - n^2/(2\Delta)^2}$ we rewrite Eq. (\ref{Eq:SolveWithSum}) as
\begin{equation}
P^D_{n}(t) = e^{-(W + \gamma)t}\theta\left(t^2 - \frac{n^2}{4\Delta^2}\right)\frac{W}{4\Delta}\sum\limits_{k = 0}^{\infty}\frac{\left(\frac{x}{2}\right)^k}{\Gamma^2\left(\frac{k}{2} + 1\right)}
\end{equation}

To evaluate the series we need to split it in two sub-series $k = 2m$ and $k = 2m + 1$. Using the definitions \cite{abramowitz1972handbook} 

\begin{equation}\label{Eq:ModBesselStruve}
\mathrm{I}_0(x) = \sum\limits_{m = 0}^{\infty}\frac{\left(\frac{x}{2}\right)^{2 m}}{\Gamma^2\left(m + 1\right)};\, \mathrm{L}_0(x) = \sum\limits_{m = 0}^{\infty}\frac{\left(\frac{x}{2}\right)^{2 m + 1}}{\Gamma^2\left(m + \frac{3}{2}\right)}
\end{equation}
we obtain the expression for the \textit{diffusive} component as 
\begin{multline}\label{Eq:PD}
P^{D}_n(t) = e^{-(W + \gamma) t}\frac{W}{4\Delta}\Bigg[\mathrm{I}_0\left(W\sqrt{t^2 - \frac{n^2}{4\Delta^2}}\right) \\
+ \mathrm{L}_0\left(W\sqrt{t^2 - \frac{n^2}{4\Delta^2}}\right)\Bigg]\theta\left(t^2 - \frac{n^2}{4\Delta^2}\right)
\end{multline}

Composing Eqs. (\ref{Eq:P0}, \ref{Eq:PD}) we obtain the final analytical solution (see Eq. (\ref{Eq:LiouvillAnswer})), which is the main result of the present study. Below we discuss different asymptotic regimes following from this result.

\section{\label{Asympt} Discussion}

To reveal how  Eq.(\ref{Eq:LiouvillAnswer}) describes \textit{ballistic} and \textit{diffusive} regimes we need to consider the asymptotic limits and discuss transitions between them. Experimentally and computationally, the energy transport time can be characterized by the dependence $T_{max}(n)$, which is the time required for the intensity on the $n$-th monomer to reach its maximum.\cite{rubtsova2013ballistic} Another interesting characteristics of energy propagation is $P_{max}$, which is the maximal intensity at site $n$ taken at the time $T_{max}$. Though our model is discreet, it is convenient to introduce spatial coordinate $x = n a$ in the asymptotic limits so that $P_n(t)\longrightarrow P(x, t)$.  We also introduce the characteristic velocity $v = 2a\Delta$, which represents the maximum group velocity of the optical phonon (see the end of Sec. \ref{GenSol}). The summary of the results is given in table \ref{Tab:Table}.  

\begin{figure}
\includegraphics[scale = .4]{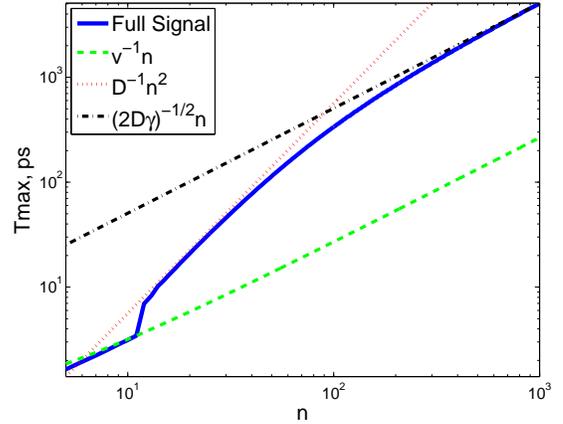}
\caption{\label{Fig:TmaxVSn} Log -- Log representation of the energy transport time $T_{max}$ as a function of distance expressed in site numbers with the key asymptotics from Eqs. (\ref{Eq:B}), (\ref{Eq:D}), (\ref{Eq:DA}), also shown in Table \ref{Tab:Table}, associated with the \textit{ballistic} (green dashed line), \textit{diffusive} (red dotted line) and \textit{directed diffusion} (black dash-dotted line) regimes, respectively. $\Delta~=~10$~cm$^{-1}$, $W~=~0.8$~ps$^{-1}$, $\gamma~=~0.001$~ps$^{-1}$}
\end{figure}

\begin{figure}
\includegraphics[scale = .4]{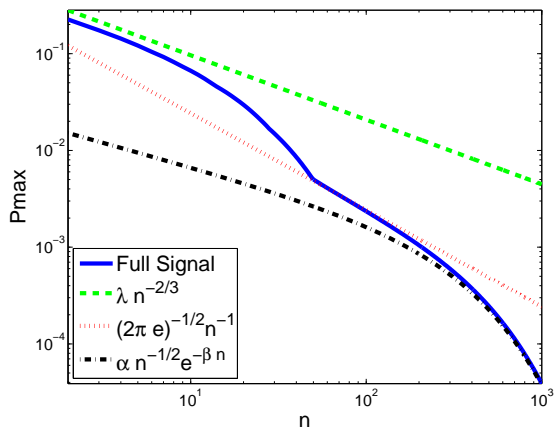}
\caption{\label{Fig:PmaxVSn} Log - Log representation of maximal intensity $P_{max}$ as a function of distance expressed in site numbers with the key asymptotics  from Eqs. (\ref{Eq:B}), (\ref{Eq:D}), (\ref{Eq:DA}), also shown in Table \ref{Tab:Table}, associated with the \textit{ballistic}(green dashed line), \textit{diffusive} (red dotted line) and \textit{directed diffusion} (black dash-dotted line) regimes respectively. $\Delta~=~10$~cm$^{-1}$, $W = 0.2$~ps$^{-1}$, $\gamma = 0.0003$~ps$^{-1}$}
\end{figure}

The \textit{ballistic} transport, Eq. (\ref{Eq:P0}), dominates at short times, $t < 1/(W + \gamma)$, where the decoherence can be neglected. For $n \gg 1$  it can be shown\cite{debye1909naherungsformeln, watson1995treatise, chishtie2008investigation, abramowitz1972handbook}  that the Bessel function $\mathrm{J}_{n}(a)$ has its first maximum at $n \approx a$ and the function amplitude $\mathrm{J}_{n}(n) \propto n^{-1/3}$. Then the energy transport time and the maximal intensity  can be estimated as a function of distance, $x$,   
\begin{equation}
\label{Eq:B}
T_{max}(x) = \frac{x}{v}; \quad  P_{max}(x)\propto  x^{-2/3}e^{-\frac{W + \gamma}{v}x} 
\end{equation}
indicating that the wavepacket in the \textit{ballistic} regime moves with the maximum group velocity (see dashed green lines in FIGs.~\ref{Fig:TmaxVSn}, \ref{Fig:PmaxVSn} and Table \ref{Tab:Table}).

The second limit, $W^{-1} < t < \gamma^{-1}$, corresponds to the \textit{diffusive} behavior ( the \textit{ballistic} component is suppressed exponentially in this regime, while the dissipation is still not significant). Using the expansion for the Bessel and Struve functions\cite{abramowitz1972handbook} the standard diffusive behavior can be reproduced as 
\begin{equation}\label{Eq:D}
T_{max}(x) = x^2/(2 D); \quad P_{max}(x) \propto x^{-1}e^{-\frac{a^2\gamma}{D}x^2}
\end{equation}
where $D = v^2 a^2/W$ (see Sec. \ref{GenSol}). One can describe the transport in this regime using the time varying instantaneous velocity $\dot{x}(t) = \sqrt{D/(2 t)}$. (red dotted lines in FIGs.~\ref{Fig:TmaxVSn}, \ref{Fig:PmaxVSn} and Table \ref{Tab:Table}). 

In the case of strong dissipation, $\gamma^{-1} < t$, the asymptotics changes and we come to the regime of "directed diffusion" affected by dissipation, where the linear dependence of the energy transport time on distance is restored
\begin{equation}\label{Eq:DA}
T_{max}(x) = \frac{x}{\sqrt{4\gamma D}}; \quad P_{max}(x) \propto  x^{-1/2}e^{-\frac{\tilde{v}}{2 D} x}
\end{equation}
and  $\tilde{v} = \sqrt{4\gamma D}$ is a new speed of the energy propagation. In this regime the straight transport is more efficient than the random walk because of the high chance of absorption for longer paths.

All three regimes are illustrated in FIGs.~\ref{Fig:TmaxVSn}, \ref{Fig:PmaxVSn} in logarithmic scale. The blue solid line  corresponds to the exact expression Eq. (\ref{Eq:LiouvillAnswer}), where we set $a = 1$, $\alpha = \left[\gamma/(4\pi^2 D) \right]^{1/4}$, $\beta = \tilde{v}/(2 D)$, $\lambda~=~(2/9)^{1/3}/\Gamma(2/3)$. The coupling was selected $\Delta~=~10$~cm$^{-1}$, corresponds to experimental data fit; \cite{rubtsova2014temperature} different $W$ and $\gamma$ parameters  were selected for convenience of illustrating clearly the transitions between the regimes (1~cm$^{-1}$~=~0.03~ps$^{-1}$).   

\begin{figure}
\includegraphics[scale = .4]{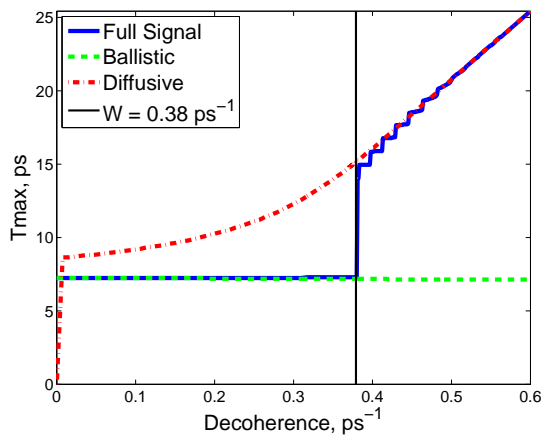}
\caption{\label{Fig:TmaxVSW}Energy transport time at the 25-th site as a function of decoherence rate $W$; $\Delta~=~10$~cm$^{-1}$, $\gamma~=~0$~ps. The sharp transition between \textit{ballistic} (green dashed line) and \textit{diffusive} (red dash-dotted line) regimes occurs at the decoherence rate $W = 0.38$~ps$^{-1}$} 
\end{figure}

It is interesting to analyze how the energy transport time depends on the decoherence rate, $W$ (FIG.~\ref{Fig:TmaxVSW}). The selection of the coupling, $\Delta = 10$~cm$^{-1}$, while somewhat arbitrary, provides qualitative agreement with the experimental data of Ref. \cite{rubtsova2014temperature}. The dissipation rate does not affect the transition between the \textit{ballistic} and \textit{diffusive} regimes, as long as $\gamma \ll W$, while at $\gamma\sim W$ the transition becomes smooth. For the sake of simplicity the results computed with $\gamma = 0$ are shown. The reported site number, $n = 25$, is chosen to satisfy the condition $n \gg 1$. One can see that the sharp transition occurs on the 25-th site at the decoherence rate $W = 0.38$~cm$^{-1}$.   

\begin{figure}
\includegraphics[scale = .4]{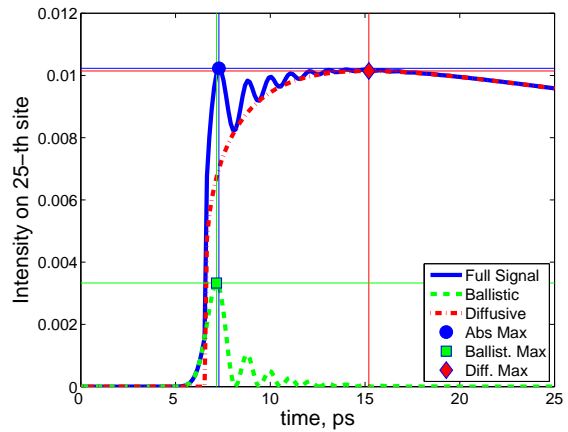}
\caption{\label{Fig:PvsT}Intensity on 25-th site as a function of time at decoherence rate $W~=~0.38$~ps$^{-1}$; $\Delta~=~10$~cm$^{-1}$, $\gamma~=~0$~ps. The full signal (blue solid line) has two equal maxima associated with the \textit{ballistic} (7~ps) and \textit{diffusive} (15~ps) components}
\end{figure}

This transition is also illustrated in  FIG.~\ref{Fig:PvsT}, where the intensity at the $25$-th site is shown as a function of time.
One can see  two maxima associated with the \textit{ballistic} (at 7~ps) and \textit{diffusive} (at 15 ps) wave fronts propagating with very different speeds. At shorter distances the \textit{ballistic} transport dominates, while at longer distances the \textit{diffusion} becomes more important (FIGs.~\ref{Fig:TmaxVSn}, \ref{Fig:PmaxVSn}). The crossover occurs therefore in discontinuous manner when the two mechanisms provide similar intensity contributions.  The signature of such crossover was observed in the temperature dependence of the  energy transport time in perfluoroalkanes.\cite{rubtsova2014temperature} 

\begin{table*}
\caption{\label{Tab:Table}\protect Characteristics of three vibrational energy transport regimes: \textit{ballistic}, \textit{diffusive} and \textit{directed diffusive}. (Note that $D = a^2\Delta/W$, $\lambda = (2/9)^{1/3}/\Gamma(2/3)$, $\alpha = \left[\gamma/(4\pi^2 D) \right]^{1/4}$)}
\begin{tabular}{|c|c|c|c|c|} \hline
\textbf{Transport mechanism}&\textbf{Time range}&\textbf{Velocity}& $\mathbf{T_{max}}$ &$\mathbf{P_{max}}$ \\ \hline

	Ballistic &$0 < t < W^{-1}$& $v = 2a\Delta$& $\frac{x}{2 a\Delta}$ &$\lambda x^{-2/3}e^{-\frac{W + \gamma}{v}x}$ \\ \hline
	
	Diffusive &$W^{-1} < t < \gamma^{-1}$ &$\dot{x}(t) = \sqrt{\frac{D}{2 t}}$ &$\frac{x^2}{2 D}$&$(2\pi e)^{{-1/2}} x^{-1}
                                                                                                    e^{-\frac{a^2\gamma}{D}x^2}$\\ \hline
 
Directed 
diffusion       &$t > \gamma^{-1}$ &$\tilde{v} = \sqrt{4\gamma D}$&$\frac{x}{\sqrt{4\gamma D}}$&$\alpha x^{-1/2}
                                                                                                   e^{-\frac{\tilde{v}}{2 D} x}$\\ \hline
\end{tabular}
\end{table*}

\section{\label{Conclusion} Conclusions}
We obtained the exact solution for the space-time represented vibrational energy \textit{ballistic} transport affected by decoherence in a quasi-continuous limit. 
We described accurately various asymptotic analytical regimes of interest, all subjects to experimental verification. 
We predict a sharp first-order-like phase transition between 	\textit{ballistic} and \textit{diffusive} transport regimes in a qualitative agreement with the recent experimental data. \cite{rubtsova2014temperature}  
Many  questions need to be addressed, including  identification of a specific vibrational mode responsible for the energy transport in a particular oligomer  and accurate analysis of decoherence and dissipation for various chain structures. 

\begin{acknowledgments}
Authors acknowledge the support from the NSF EPSCoR LA-SIGMA (EPS-1003897), NSF CHE-1012371, Army Research Office (vv911NF-13-1-0186) and Louisiana Board of Regents LINK (NSF(2014)-LINK-90) programs. Authors also acknowledge Abraham Nitzan and Andrii Maksymov for fruitful suggestions.
\end{acknowledgments}
\newpage
\bibliography{Reference}
\end{document}